\documentstyle[12pt]{JHEP}

\newcommand{\ie}{{\em i.e.~}}
\newcommand{\eg}{{\em e.g.~}}
\newcommand{\alpr}{{\alpha^{\prime}}}

\newcommand{\be}{\begin{equation}}
\newcommand{\ee}{\end{equation}}
\newcommand{\ba}{\begin{eqnarray}}
\newcommand{\ea}{\end{eqnarray}}

\title{RG-flows and Open/Closed String Duality } 
\author{Massimo Bianchi and Jose F. Morales \\ 
{\it Dipartimento di Fisica, \ Universit{\`a} di Roma} \  
{\it ``Tor Vergata''\\ I.N.F.N.\ -\ Sezione di Roma} \ 
{\it ``Tor Vergata''}\\ {\it Via della Ricerca  Scientifica, 1} 
\\ {\it 00173 \ Roma, \ ITALY} }

\abstract{We discuss the interpaly between IR and UV divergences in 
theories with open and unoriented strings in view of the AdS/CFT 
correspondence. We start by deriving
general formulas for the computation of threshold corrections to 
gauge couplings in generic configurations with open and unoriented strings.
These allow us to discuss the IR/UV correspondence between 
beta-function coefficients and ``dilaton'' tadpoles for several brane
configurations probed by D3-branes.
Finally we comment on the AdS supergravity 
descriptions of gauge theories that are (super)conformal 
in the large $N$ limit.}

\preprint{ROM2F-2000-33}
\begin{document}
\setcounter{page}{1}

\section{Introduction}
The interplay between open and closed string divergences in amplitudes 
around non-trivial configurations with Dirichlet branes and orientifold
planes is of particular import
to all recent developments in string theory. 

On the one hand, it has been recently shown in \cite{bm} 
that on very general grounds closed string divergences 
associated to massless R-R fields disappearing 
into the vacuum (R-R tadpoles) are always associated to anomalies in the 
open-string theory that describe the excitations of the brane background.
On the other hand, NS-NS tadpoles, 
such as the dilaton tadpole, do not necessarily imply a 
pathology of the theory rather they signal an instability of the vacuum 
configuration one is expanding around. Long time ago Fischler and Susskind 
have proposed an algorithm for systematically reabsorbing such divergences
\cite{fs}. More recently the role of the dilaton tadpole has been analyzed in 
connection to the AdS/CFT correspondence \cite{aa}. Interestingly the 
dilaton tadpole on the disk and projective plane
for some non-tachyonic unoriented projections \cite{ausag} of the type OB theory
have been shown to be in quantitative agreement with the one-loop
$\beta$-function coefficient of the four dimensional gauge theory 
living on D3-brane probes \cite{aa, bfl}. 
Given the standard relation between ten dimensional dilaton and gauge 
coupling on the brane
it is natural to ask to what extent the relation found in \cite{aa} 
generalizes to other configurations with open and unoriented strings.

In this respect, it is convienient to observe 
that up to some ``innocuous'' T-duality transformation, 
vacuum configurations with open and unoriented 
strings can be related to 
F-theory backgrounds or more efficiently  
configurations with D7-branes and O7-planes  
wrapping possibly non-trivial cycles.
Most of the emphasis so far has been on the configurations with constant 
dilaton that can pe probed by D3-branes giving rise to interesting 
(super)conformal field theories \cite{afm}. Non-conformal gauge theories
arising on D3-brane probes of tachyonic type O backgrounds have been 
studied \cite{kt,minahan}. Some RG-flows obtained by perturbing 
superconformal fixed points by relevant operators have been studied 
by various groups \cite{gppz, fgpw, dz}. Although the five-dimensional 
perspective allows one to derive a holographic c-theorem and in principle to 
compute mass spectra and correlation functions in the boundary gauge 
theory \cite{dwf} a lift to ten dimensions seems to be 
required in order to understand the fate of some curvature 
singularities \cite{jpp} and the running of the gauge coupling \cite{pw}. 
Related work on the subject also appeared in \cite{eacg}.

Our emphasis will be on theories 
that arise from a (super-)conformal background by displacing 
some of the branes in such a way that 
the 10-D dilaton tadpole does not vanish locally.
Following \cite{aa} we expect that the resulting gauge theory on the brane 
should display a running coupling with a one-loop coefficient related to the 
dilaton tadpole on the disk/projective plane. 
We will find that the quantitative agreement between dilaton tadpole and 
$\beta$-function coefficient can be substantiated for sectors 
of the open-string spectrum that enjoy   
${\cal N}=2$ supersymmetry (\ie 8 supercharges). 
The resulting theory need not necessarily be 
supersymmetric because the supercharges preserved in one sector can be 
different from the supercharges preserved in another sector. 
 We will argue that this underlying 
``fictitious'' ${\cal N}=2$ supersymmetry protects the IR/UV correspondence
between tadpoles and RG-flows from higher loop corrections at the
leading order in the 't Hooft coupling $\hat{g}^2=g^2 N$ 
for all ${\cal N}=1,2$ examples presented here.    
This is not the case for the ${\cal N}=0$ examples presented in \cite{aa} for
which the correspondence still works at the 
one-loop level but is expected to receive non-trivial corrections at 
higher orders both in the SYM and in the Supergravity descriptions. 
A quantitative relation between ``properly regularized''
dilaton tadpole at one-loop (annulus, ....) and two-loop coefficient of the 
$\beta$ function would be of particular significance but very hard to establish 
because of the 
well-known ambiguities in superstring perturbation theory beyond one-loop.

The plan of the paper is as follows.
In Section 2 we discuss divergences in string amplitudes in the presence of 
non-trivial backgrounds for the boundary (gauge) fields.
Closely following \cite{bf,abd}, we derive
general formulas for the computation of threshold corrections to 
gauge couplings in configurations with open and unoriented strings.
The results apply to any configuration of open strings 
satisfying the basic consistency requirements developed in \cite{aug,bs}.
In Section 3 we discuss the IR/UV correspondence between 
$\beta$-function coefficients and 
``dilaton'' tadpoles for several examples of brane
configurations constructed in terms of D3-branes, D7-branes and orientifold 
planes.
In Section 4 we comment on the AdS supergravity 
descriptions of gauge theories that are (super)conformal 
in the large $N$ limit. In particular we show how the subleading logarithmic 
running of the gauge coupling is quantitatively reproduced in the 
holographic correspondence by the evolution of the 10-D dilaton 
caused by the presence of non-vanishing tadpoles on the disk and 
projective plane.

\section{Infrared/Ultraviolet divergences in open string amplitudes}

In this section we apply the techniques developed in \cite{bf,abd} to 
the study of the infrared/ultraviolet divergences of one-loop
open string partition functions in the presence of non-trivial 
backgrounds for the boundary gauge fields.
  
String amplitudes of this kind are efficiently encoded
in helicity supertraces over the Hilbert space of open
string states connecting different boundaries and crosscaps.
The Hilbert space of open string states can be 
decomposed into a (generically infinite) 
sum ${\cal H} = \oplus_{i} {\cal H}_i$ of Verma moduli
labeled by the primary fields $\Phi^i$
of some chiral algebra ${\cal G}$, that includes  
an ${\cal N}=1$ superconformal algebra at least.
The subspace ${\cal H}_i$ consists in the tower of 
descendants under ${\cal G}$ of the groundstate $|\Delta_{i}\rangle = 
\Phi^i|0\rangle$. 
The helicity supertrace over states in ${\cal H}_i$ 
is defined by the holomorphic elliptic function
\be
{\cal X}_{i}(u|\tau) = {\rm STr}_{{\cal H}_i}^{\prime}\, 
\left[q^{L_{0}-{c\over24}} e^{2 \pi i u_{\mu\nu} J^{\mu\nu}}\right]
\quad , 
\label{character}
\ee  
with $L_{0}$ the Virasoro generator (open string Hamiltonian),
$J^{\mu\nu}$ the generators of the $SO(D-2)$ (little) Lorentz group and 
$q=e^{2\pi i \tau}$. 
A prime in (\ref{character}) indicates the 
omission of the bosonic zero-modes contribution to the trace, that will
be explicitly displayed in the next formulas.
For ``rectangular'' toroidal 
compactifications, \ie for vanishing off-diagonal components of the metric
and B-field, it is given by
\be
{1 \over t^{D\over 2}}\, \Sigma_\Lambda(t) \equiv 
{1 \over t^{D\over 2}}\,P_{n_\Lambda}(t) W_{d_\Lambda}(t) =
{1 \over t^{D\over 2}}\,\sum_{(m_i,n_j)\in {\bf Z}^{n_\Lambda+d_\Lambda}}\, 
e^{-\pi t\left({\alpha^\prime m_i^2\over R_i^2}+
{n_j^2 R_j^2 \over \alpha^\prime}\right)}
\label{lattice}
\ee
where $D$ is the number of non-compact directions\footnote{The 
resulting hyperplane 
is to be identified with the spacetime under consideration. We assume
NN boundary conditions along it for all open string
excitations of the brane configuration.},
$P_{n_\Lambda}(t)$($W_{d_\Lambda}(t)$) are $n_\Lambda$
($d_\Lambda$)-dimensional 
momentum(winding) lattice  
sums for open strings with NN(DD) 
boundary conditions labeled by the lattice $\Lambda$ that defines the 
torus $T^{n_\Lambda+d_\Lambda} = R^{n_\Lambda+d_\Lambda}/ \Lambda$.   

Modular invariance (of the bulk theory) ensures that
helicity supertraces (\ref{chis}) can be written as sums 
over even spin structures\footnote{The odd spin structure
contribution, which is crucial in establishing the relation between 
R-R tadpoles and anomalies, will play no role in the thresholds we consider
hereafter.} 
\be
{\cal X}_{i}(u|\tau) = \sum_{(\alpha,\beta) \neq ({1\over 2},{1\over 2})} 
{\cal X}_{i}{\alpha\brack\beta}(\tau) 
{\cal F}{\alpha\brack\beta}(u|\tau)=h^i(u)+O(q)
\label{chis}
\ee 
with ${\cal X}_{i}{\alpha\brack\beta}(\tau)$ the contribution
of the sector with $(\alpha,\beta)$ spin structure to the open string
partition function and
\be
{\cal F}{\alpha\brack\beta}(u|\tau)=\prod_{m=1}^{{D-2 \over 2}}
{\vartheta{\alpha\brack \beta}(u_m|\tau) \over
\vartheta{\alpha\brack \beta}(0|\tau)}\,
{\vartheta{{1\over 2}\brack {1\over 2}}^\prime(0|\tau) \over
\vartheta{{1\over 2}\brack {1\over 2}}(u_m|\tau)}\,u_m\quad 
\ee
is the spacetime helicity supertrace which has been normalized 
to one for $u=0$.
Without loss of generality $u_{\mu\nu}$ has been taken  
in the block diagonal form $u=u_m \delta_{2m,2m-1}$.
In the right hand side of (\ref{chis}) we have assumed that the theories
we deal with are free of tachyons. For
later convenience we have introduced 
the notation $h^i(u)$ for the helicity supertrace
restricted to massless states in ${\cal H}_i$. It can be written as
\be
h_i(u)=\sum_r h_i^{(r)} u^r={\rm Str}_{{\cal H}_i}^{\rm massless}
\prod_{m}\left[{\pi u_m \over sin(\pi u_m)}\right]
\sum_m cos(2 \pi u_m J_m)
\label{h}
\ee
with $J_m$ the eigenvalue of the helicity operator in the $(2m,2m-1)$ plane.

The modular transformation properties of the character-valued
partition functions ${\cal X}_{i}(u|\tau)$
are determined by the modular transformation properties
of the characters. Denoting by 
$S_i^j$ and $T^j_i$ 
the matrices representing the generators of the modular group 
in the ${\cal X}_{i}(0|\tau)$ basis, one 
finds   
\ba
T: \qquad {\cal X}_i(u|\tau+1) 
&=& e^{2\pi i (\Delta_{i}-c/24)}
{\cal X}_i(u|\tau)
\nonumber\\
S: \qquad {\cal X}_i\left({u\over \tau}|-{1 \over\tau}\right)
&=& (i \tau)^{-\frac{D+n+d-2}{2}}S^j_i
{\cal X}_j (u|\tau) \quad
\label{modular}
\ea
where $\Delta_i$ is the conformal dimension of the primary field $\Phi_i$.
Hereafter a sum over repeated indices is always 
understood: $a^i b^i c^i...\equiv \sum_i \,a^i b^i c^i...$.  

Following \cite{bp,aclny} one can write the open string
partition function in the presence of a constant electromagnetic
background
$F_{12}=f\, Q$, with $Q$ a generator in the Cartan-subalgebra  
normalized in such a way that $tr_{fund}\, Q^2={1\over 2}$. The net result 
of the non-trivial background is a shift of the oscillator 
frequencies of the complex coordinate $X^1+i X^2$ by an amount 
$\epsilon$ given by
\be
\pi \epsilon_{ab}={\rm arctan}(\pi f q_a)+{\rm arctan}(\pi f q_b)
\ee
with $q_a,q_b$ the charges of the boundaries $a,b$ respectively 
where the open string
ends (for the M\"obius strip $q_a=q_b$).
Generalizing \cite{bf,abd}, 
the (direct channel) partition function for the open string excitations of a 
brane configuration in the presence of 
an electromagnetic background can be 
written as
\ba
{\cal A}(f)&=&
{1\over 2}\, 
A^{i\Lambda}_{ab} 
{\rm Tr}_{n^{a}\otimes n^{b}} {\pi f(q_a+q_b)\over \epsilon_{ab}}\,
\int
{dt\over t^{1+{D\over 2}}}\,  
{\cal X}_{i} \left(\epsilon_{ab}{it\over 2}|{it\over 2}\right)
\Sigma_\Lambda(t)
\nonumber\\
{\cal M}(f)&=&
{1\over 2}\, M^{i\Lambda}_{a} {\rm Tr}_{n^{a}}
\,{2\pi f q_a\over \epsilon_{aa}}\, 
\int
{dt\over t^{1+{D\over 2}}} 
 \widehat{\cal X}_{i}
\left(\epsilon_{aa}{it\over 2}|
{it\over 2} + {1\over 2}\right)\Sigma_\Lambda(t)  
\label{direct} 
\ea     
with the integers $A^{i\Lambda}_{ab}$ and $M^{i\Lambda}_{a}$ 
counting the number of times the open string sector $i$ with
internal momenta belonging to the lattice $\Lambda$ and ends 
of type $(a,b)$ and $(a,a)$ runs along 
the Annulus and M\"obius-strip loop, respectively.   
Alternatively, after the  
change of variables $\ell_A ={2\over t}$ and $\ell_M={1\over 2t}$
in (\ref{direct}), one can reinterpret these amplitudes 
as tree level exchange (transverse channel) of closed string states
between the boundaries and crosscaps 
\ba
{\cal A}(f) &=&
{2\over 2}^{-{D+n+d\over 2}}  
B^{i\Lambda}_{a}  B^{i\Lambda}_{b}{\rm Tr}_{n^{a}\otimes n^{b}}
{\pi f(q_a +q_b)\over \epsilon_{ab}}\,\int d\ell \, 
{\cal X} _{i}(\epsilon_{ab}|i\ell )\Sigma_{\Lambda}({\ell\over 2})\nonumber\\
{\cal M}(f) &=&{2\over 2}  
\Gamma^{i\Lambda} B^{i\Lambda}_{a}{\rm Tr}_{n^{a}}
{2 \pi f q_a\over \epsilon_{aa}}\,\int d\ell\,  
\widehat{\cal X}_{i}({\epsilon_{aa}\over 2}|i\ell +{1 \over 2})\,
\Sigma^{{\rm e}}_{\Lambda}({\ell\over 2}) \quad .
\label{transverse} 
\ea 
$\Gamma^{i\Lambda}$ and $B^{i\Lambda}_{a}$ are
the reflection coefficients (one-point function) of a state in the
character ``$i$'' with internal momentum in $\Lambda$ in front of a 
crosscap or a boundary of type $a$, respectively.
They are related to the integer coefficients in the direct channel 
by suitable modular transformations:
\ba
A^{i\Lambda}_{ab} &=&  
S_{i^\prime}^{i}s_{\Lambda^\prime}^{\Lambda} B^{i^\prime \Lambda^\prime}_{a}
B^{i^\prime \Lambda^\prime}_{b} \nonumber\\
M^{i\Lambda}_{a} &=&  
P^{i}_{i^\prime} s_{\Lambda^\prime}^{\Lambda} 
\Gamma^{i^\prime \Lambda^\prime} 
B^{i^\prime \Lambda^\prime}_{a}
\label{worldsheet}
\ea
with $S_{i^\prime}^{i}$ defined in (\ref{modular}), 
$P \equiv T^{1/2} S T^2 S T^{1/2}$ and 
$s_{\Lambda^\prime}^{\Lambda}$ the result of the Poisson resummation
\be
\Sigma_\Lambda\left({1\over t}\right)=P_{n_\Lambda}\left({1\over t}\right)
W_{d_\Lambda}\left({1\over t}\right) 
=t^{-{n_\Lambda+d_\Lambda\over 2}}
s_{\Lambda^\prime}^{\Lambda}\Sigma_{\Lambda^\prime}(t)=
t^{-{n_\Lambda+d_\Lambda\over 2}}{v_{n_\Lambda}\over 
v_{d_\Lambda}} W_{n_\Lambda}(t) P_{d_\Lambda}(t)
\label{poisson}
\ee
$v_{n_\Lambda}, v_{d_\Lambda}$ are the volumes of the NN, DD compact 
hyperplanes (labeled by $\Lambda$) measured in units of $2\pi\alpha^\prime$.  
Finally the notation $\Sigma^{{\rm e}}_{\Lambda}(\ell )$ in the 
transverse M\"obius amplitude denotes a restriction 
to even integers in the lattice sum, \ie to  
the states that can be reflected by a crosscap \cite{ps, gp}.

One-loop open string amplitudes involving an arbitrary number of 
boundary field insertions $F_{12}$ can be derived from (\ref{direct}) by 
expanding to a suitable order in the field strength background $f$. 
In general they suffer from both infrared 
and ultraviolet divergences. Introducing in (\ref{direct},\ref{transverse}) 
$\lambda_{IR}$($\lambda_{UV}$) cutoff's for the infrared(ultraviolet) 
divergences
one is left with the regularized modular integrals
\be
{\cal A}(f)+{\cal M}(f)=\int_{2\lambda_{UV}}^{2\lambda_{IR}} 
{dt\over t} A(f|t)+
\int_{\lambda_{UV}\over 2}^{\lambda_{IR}\over 2}{dt\over t}
M(f|t) =\int^{\lambda_{UV}^{-1}}_{\lambda_{IR}^{-1}}
d\ell\left[\widetilde{A}(f|\ell)+\widetilde{M}(f|\ell)\right].
\label{cutoff}
\ee
Following \cite{bf,abd} we have choosen to work with a uniform cutoff
in the transverse channel that give rise to non-uniform cutoffs in 
the direct channel. 

The (divergent) coupling constants of the operators 
${\rm Tr}_{{\cal R}}{\cal F}^r$ 
(${\cal R}$ running over the spectrum of 
open string representations)
are compactly encoded in the 
coefficients of the expansion in powers of $f$ of ``generating 
$\beta^a(f)$ functions'' defined by
\ba
\beta_{IR}(f) &\equiv&
\sum_r \beta_{IR}^{a(r)}{\rm Tr}_{n_a}(\pi f q_a)^r
\equiv \lambda_{IR}{\partial\over \partial \lambda_{IR}}\left[
\int_{2\lambda_{UV}}^{2\lambda_{IR}}
{dt\over t}\, A(f|t)
+\int_{\lambda_{UV}\over 2}^{\lambda_{IR}\over 2}{dt\over t}\,
M(f|t)\right]\nonumber\\
&&=\lim_{\lambda_{IR}\rightarrow\infty}\left[A(f|2 \lambda_{IR})+
M(f|{\lambda_{IR}\over 2})\right] 
\label{betair}\\ 
\beta_{UV}(f)&\equiv&
\sum_r  \beta_{UV}^{a(r)}{\rm Tr}_{n_a}(\pi f q_a)^r
\equiv -\lambda_{UV} {\partial \over \partial \lambda_{UV}}
\int^{\lambda_{UV}^{-1}}_{\lambda_{IR}^{-1}}
{d\ell}\, \left[\widetilde{A}(f|\ell)+\widetilde{M}(f|\ell)\right]
\nonumber\\
&&=\lim_{\lambda_{UV}\rightarrow 0}\, \lambda_{UV}^{-1}\,\left[
\widetilde{A}(f|\lambda_{UV}^{-1})+
\widetilde{M}(f|\lambda_{UV}^{-1})\right] \quad .
\label{betauv}
\ea
Infrared divergences in (\ref{betair}) are associated to 
the running of the gauge couplings in the effective gauge theory
living on the branes. Ultraviolet divergences arise 
from the factorization of the transverse amplitudes into  
the disk and projective plane and are associated to the 
infinite time propagation of 
massless supergravity states. Both kinds of divergences receive 
only contributions from massless states in the relevant description
and therefore can be studied using an effective Supergravity (bulk theory)
or SYM (brane theory) picture according to the case.
The answers disagree in general but under special 
circumstances they lead to the same result. The main aim 
of this paper is to explore branches in the moduli space of certain brane
configurations where RG-flows in the ``boundary'' gauge theory receive
an accurate description in terms of the  ``bulk'' supergravity theory.

We will concentrate our attention in RG-flows of the irreducible couplings 
${\rm Tr}_{n_a} {\cal F}^r$ in the gauge theory.  
Plugging (\ref{direct}) in (\ref{betair}) and noticing that 
in the limit $t\rightarrow\infty$ the leading  contributions
(at a given order $r$) comes from the expansion of ${\cal X}_i$ in
its first argument, one can simply replace 
$\epsilon_{ab}$ with $\pi f(q_a+q_b)$. One is then left  
with the $\beta$-function coefficients  
\be
\beta_{IR}^{a(r)}={h_i^{(r)}\over 2}\,
\left[2^{1-{D\over 2}}\,A^{i\Lambda}_{ab} n^{b}\Sigma_\Lambda(2\lambda_{IR}) 
+2^{{D\over 2}-r}M^{i\Lambda}_a  \Sigma_\Lambda({\lambda_{IR}\over 2}) \right]
\lambda_{IR}^{r-{D\over 2}}
\label{betairh}
\ee
with $h_i^{(r)}$ the massless helicity supertrace coefficients
defined through (\ref{h}). 
In particular, after expanding (\ref{h}) one gets for the first 
few coefficients
\be
h_i(u)=\sum_r h_i^{(r)} {u^r}
={\rm Str}_i\, 1+{u^2} {\rm Str}_i(-2J^2+{1\over 6})+
{u^4\over 3} {\rm Str}_i(2J^4-J^2+{7 \over 120})+ ...
\label{coefb}
\ee
with $J=J_{12}$ the (eigenvalue of the)  helicity operator in the $(12)$-plane.
Evaluating the helicity traces (\ref{coefb}) yields 
\ba
h_i^{(0)} &=& 2 n^i_V+n^i_O-(n^i_S+n^i_C)\nonumber\\
h_i^{(2)} &=& -{11\over 3} n^i_V+{1\over 6}n^i_O+{2\over 3}(n^i_S+n^i_C)
\nonumber\\ 
h_i^{(4)} &=& {1\over 360}\left[
254 n^i_V+7 n^i_O+16 (n^i_S+n^i_C)\right]
\label{betaD}
\ea
where $n^i_O$, $n^i_V$, $n^i_S$ and $n^i_C$ are integers counting  
the number of massless states in the character ``$i$''
transforming as scalars, vectors, left- 
and right-handed spinors, respectively, 
\ie having helicity $J=0$, $J=\pm 1$,  
and $J=\pm {1\over 2}$.
One recognizes in $h_i^{(2)}$ the familiar $\beta$-function coefficients 
associated to the running of the ${\cal F}^2$ term in four dimensional
gauge theories. 
To this end one should recall that $n_S^i=n_C^i=n_F^i/2$ 
as a consequence of the CPT theorem in $D=4$.
Much in the same way, $h_i^{(0)}$ and $h_i^{(4)}$ can be associated
to the logarithmically running self-energy (0-point function) in $D=0$
and four-point coupling ${\rm Tr}_{{\cal R}} {\cal F}^4$        
in $D=8$.

A similar analysis can be performed in the study of 
ultraviolet divergences
of open string partition functions in an electromagnetic 
background. 
After some simple trigonometric manipulations one finds
\footnote{The discrepancies in signs with respect to \cite{gg} is due
to our choice (as in \cite{bf}) of defining the charges at the
two ends of the string without a relative minus sign.}  
\ba
\widetilde{h}_i(f)&=&
h_i(\epsilon_{ab}){\pi f(q_a+q_b)\over \epsilon_{ab}}\nonumber\\
&&={\pi f(q_a+q_b)\over sin(\pi\epsilon_{ab})}\left[
n^i_0+2 cos(2\pi\epsilon_{ab}) n^i_V-cos(\pi\epsilon_{ab})
(n^i_S+n^i_C)\right]
\nonumber\\
&=&{1\over\sqrt{1+\pi^2 f^2 q_a^2}\sqrt{1+\pi^2 f^2 q_b^2}}
\left[(n^i_0+2 n^i_V)(1+\pi^2 f^2 q_a^2)(1+\pi^2 f^2 q_b^2)\right.
\nonumber\\
&&\left. -4 n^i_V\pi^2 f^2 (q_a+q_b)^2\right]
-(n^i_S+n^i_C)(1-q_a q_b\pi^2 f^2)
\label{hbi}
\ea
By inspection of (\ref{transverse}) one sees that Annulus and
M\"obius amplitudes factorize into two disks or a disk and a crosscap
with certain number of insertions of gauge fields on the boundaries.
It is remarkable that for any consistent open-string descendant
(even without a clear geometrical meaning) 
one can deduce generalized Born-Infeld and WZ couplings much in the 
same way as in \cite{gg}. From the factorization of (\ref{hbi})
one can extract the coupling of the brane theory to the bulk 
NS-NS scalars ($n_O$), graviton and dilaton ($n_{V}$) 
and R-R tensor fields ($n_S,n_C$). 
For flat D-branes this amounts to the familiar fact that (\ref{hbi})
can be derived from the effective worldvolume action 
\be
S_{brane}=S_{BI}+S_{WZ}
\ee
with
\ba
S_{BI}&=&\int d^{p+1} e^{-\phi}\sqrt{{\rm det} ({\cal G}+{\cal F})}
\nonumber\\
&&={e^{-\phi_0}\over 4}\sqrt{1+\pi^2 f^2 q^2}\left(1+{1\over 4} h_{\mu\mu}
-{1\over 4} h_{ii}+{2\pi f b_{12}-\pi^2 f^2 q^2 (h_{11}+h_{22})\over 2(1+\pi^2 f^2 q^2)}+...
\right)
\nonumber\\
S_{WZ}&=&\int\,\left( C_{p+1}+C_{p-1}\wedge {\cal F}\right)
\label{bi}
\ea
with $h^{\mu\mu}$, $h_{ii}$ the metric components 
along and transverse to the brane respectively, $b_{12}$ the NS-NS
$B$-field fluctuation, $F_{12}=\pi q f$ and
$\phi={1\over 4}(h^\alpha_\alpha+h_i^i)$ the ten dimensional dilaton. 
The dots in (\ref{bi}) represent higher order powers of
the fluctuations
$h_{MN}$, $b_{MN}$ in the expansions $G_{MN}=\eta_{MN}+h_{MN}$,
$B_{MN}= b_{MN}$ around the flat metric $\eta_{MN}$ and zero B-field. 

Once again specializing to irreducible couplings one is left with
\be
\beta^{a(r)}_{UV}= B^{i\Lambda}_a 
\widetilde{h}_i^{(r)} 
\left[2^{-{D+n_\Lambda+d_\Lambda\over 2}}B^{i\Lambda}_{b} n^{b}\,
\Sigma_\Lambda({\lambda_{UV}^{-1}\over 2})
+\Gamma^{i\Lambda}\,\Sigma^{{\rm e}}_\Lambda({\lambda_{UV}^{-1}\over 2})
\right]\lambda_{UV}^{-1}
\label{betauvh}
\ee
As expected the ultraviolet divergences of irreducible amplitudes  
are given by a linear combination of the tadpoles
$T^{i\Lambda}=2^{-{D+n_\Lambda+d_\Lambda\over 2}}B^{i\Lambda}_{b} n^{b} +
\Gamma^{i\Lambda}$ and in a very natural fashion are uniformly cut
off in the transverse channel as suggested in \cite{abd}.

\subsection{``BPS saturated'' string amplitudes}

In the following we will consider explicit brane realizations 
for which the effective gauge theory (\ref{betairh}) and 
supergravity (\ref{betauvh}) answers for the $\beta$-function
coefficients associated to logarithmically running couplings coincide.  
We will consider couplings in the $D$-dimensional theory 
of a finite set of $N$ D-branes probing a background
geometry involving generalized 
D-branes and O-planes.
Compact directions for strings with at least one end in
the probes can have either DD or ND boundary conditions and
therefore $n_{\Lambda}=0$ for all the relevant string excitations.
In general, the $\beta$-function coefficients (\ref{betair}) and (\ref{betauv})
receive  contributions from infinite
towers of open string winding modes or KK-supergravity 
states, respectively, coming from the reduction on $T^{d}$
of a higher dimensional gauge or supergravity theory.
In order to make contact with more familiar $D$-dimensional field theory
results we will consider the decompactification limit 
$R_j>>\sqrt{\alpha^\prime}$ where the whole tower of open
string winding modes decouples 
($\lim_{R_j\rightarrow \infty} W_d(t)=1$)
and KK supergravity contributions resum to 
\ba
\lim_{R_j\rightarrow \infty} {P_d\left({\ell\over 2}\right)\over v_d}&=& 
\left({\ell\over 2}\right)^{-{d\over 2}}
\nonumber\\ 
 \lim_{R_j\rightarrow \infty} {P^{e}_d\left({\ell\over 2}\right)
\over v_d}&=& 2^{-d}
 \left({\ell\over 2}\right)^{-{d\over 2}} 
\label{rinfinity}
\ea
The factor $2^{-d}$ reflects the fact that charge
and mass of the higher dimensional O-planes 
are democratically distributed between the $2^d$
lower dimensional O-planes. 
Logarithmically running couplings in both the direct and transverse
channel then require $d=2$ with $r={D \over 2}$ as the number of 
boundary insertions. The logarithmic divergences 
in the open string channel arises as usual from loop integrations 
of massless gauge particles, while in the supergravity picture 
they are associated to two-dimensional propagators of 
massless closed string states.   
The question one may raise is whether the coefficients in front of
the two logarithmic divergences coincide.
We will only consider the case with $r$ even \eg $D=0,4,8$.
One can then see that ``minimal'' supersymmetry
in these situations is enough to ensures the
agreement between the two $\beta$-function coefficients
(\ref{betair},\ref{betauv}).   
Indeed by assuming the minimal ${\cal N}=1$ supersymmetry in $D+2$  
dimensions (or equivalently
${\cal N}=2$ in $D$ dimensions) the $r={D\over 2}$-point
function belongs to the class of ``BPS 
saturated'' couplings for which 
the contributions coming from massive worldsheet  bosonic and 
fermionic states cancel against each other leaving a $t$-independent
result.

In order to see this let us consider D3-branes probes (${\cal F}^2$-term
in $D=4$) of $Z_N$ orientifold backgrounds. The string partition 
functions can be written in terms of $Z_N$ characters $\chi_{gh}$
defined in terms of the chiral amplitudes $\rho_{gh}$ through
\ba
\chi_{gh}(u)={1\over N}\sum_{l=0}^{N-1}\,
w^{l h} \rho_{g l}(u)\Sigma_{gl}={\rm STr}_{\rm g-twisted}
\left[\sum_{l=0}^{N-1}\,w^{l h} \Theta^l\,
q^{L_{0}-{c\over24}} e^{2 \pi i u_{\mu\nu} J^{\mu\nu}}\right]
\label{zncharacter}
\ea
where $\Theta$ is the $Z_N$ generator,  $g,h=0,1,...N-1$, 
$\omega=e^{2\pi i\over N}$ and
$\Sigma_{gl}$ is the lattice sum in the hyperplane that is fixed under both
$g$ and $l$ twists in the $\sigma$ and $\tau$ directions respectively .
Expanding (in powers of $u$), the helicity supertraces 
$\rho_{gh}$ in (\ref{zncharacter})
for $g,h$, $Z_N$ rotations of a $T^4$ torus one can easily check
the  $t$-independence of the leading orders in the $u$-expansions of
(\ref{zncharacter}) (`` BPS saturated term'')
\footnote{We borrow the notation and modular transformation
properties defined in the appendix A of \cite{rho}.} 
\ba
\rho_{00}(u)&=&2\pi u\,\sum_{\alpha,\beta \neq 0} 
\eta_{\alpha\beta}{\vartheta{\alpha\brack \beta}(u|\tau)\over
\vartheta{{1\over 2}\brack {1\over 2}}(u|\tau) \eta^9} 
=2\pi u\,{\vartheta{{1\over 2}\brack {1\over 2}}^{4}({u\over 2}|\tau)\over
\vartheta{{1\over 2}\brack {1\over 2}}(u|\tau)\eta^{9}} 
= u^{4}+ ....\nonumber\\
\rho_{0h}(u)&=&-8\pi u sin^2{\pi h \over N}\sum_{\alpha,\beta \neq 0}
\eta_{\alpha\beta} 
{\vartheta{\alpha\brack \beta}(u|\tau)\over
\vartheta{{1\over 2}\brack {1\over 2}}(u|\tau)} 
{\vartheta{\alpha\brack \beta}(0|\tau)\over
\eta^3}
\prod_{e=\pm}{\vartheta{\alpha\brack \beta+e h}(0|\tau)\over
\vartheta{{1\over 2}\brack {1\over 2}+e h}(0|\tau)}
\nonumber\\
&&=-8\pi u sin^2 {\pi h \over N}
{\vartheta{{1\over 2}\brack {1\over 2}}^{2}({u\over 2}|\tau)\over
\vartheta{{1\over 2}\brack {1\over 2}}(u|\tau)\eta^{3}} 
\prod_{e=\pm}{\vartheta{{1\over 2}\brack {1\over 2}-e h}({u\over 2}|\tau)\over
\vartheta{{1\over 2}\brack {1\over 2}+e h}(0|\tau)}
= -4 sin^2{\pi h \over N} u^{2}+ ....\nonumber\\
\rho_{gh}(u)&=&2\pi u\,\sum_{\alpha,\beta \neq 0}\eta_{\alpha\beta} 
{\vartheta{\alpha\brack \beta}(u|\tau)\over
\vartheta{{1\over 2}\brack {1\over 2}}(u|\tau)} 
{\vartheta{\alpha\brack \beta}(0|\tau)\over
\eta^3}
\prod_{e=\pm}{\vartheta{\alpha+e g\brack \beta+e h}(0|\tau)\over
\vartheta{{1\over 2}+e g\brack {1\over 2}+e h}(0|\tau)}
\nonumber\\
&&=2\pi u\,{\vartheta{{1\over 2}\brack {1\over 2}}^{2}({u\over 2}|\tau)\over
\vartheta{{1\over 2}\brack {1\over 2}}(u|\tau)\eta^{3}} 
\prod_{e=\pm}{\vartheta{{1\over 2}-e g\brack {1\over 2}-e h}({u\over 2}|\tau)
\over\vartheta{{1\over 2}+e g\brack {1\over 2}+e h}(0|\tau)}
=u^{2}+ ....
\label{spinsum}
\ea 
where dots stand for subleading term in the $u$-expansion. 
The $t$-independence of the leading order terms in
these expansions has already been 
appreciated in the study of spin effects associated to the interactions of
Dp and Dp+4 branes
and of anomalous couplings \cite{mss}. 
The result for the $u^{2}$ term in (\ref{spinsum}) is then
exact, admitting equivalent SYM ($t\rightarrow \infty$)
and Supergravity ($t\rightarrow 0$) descriptions, obtained 
by replacing 
$(\rho_{00},\rho_{0h},\rho_{gh})$ with $(0,-4 sin^2{\pi h\over N},1)$
in the direct and transverse string amplitudes (\ref{direct},\ref{transverse}).
In the decompactification limit (\ref{rinfinity}), the $\beta$-function
coefficients can then be written in the two alternative forms 
\ba
\beta^{(2)a}_{IR}&=&
{1\over 4}(A^{i\Lambda}_{ab} n^{b} +2 M^{i\Lambda}_a)h_i^{(2)}=
{1\over 2}{\rm l}_2({\cal R}^{a}_i)h_i^{(2)}
\nonumber\\
\beta_{UV}^{(2)a}&=&2
(2^{-3}B^{i\Lambda}_{b} n^{b} +{1\over 4}\Gamma^{i\Lambda})B^{i\Lambda}_a 
h_i^{(2)}=2\hat{T}^{i\Lambda} B^{i\Lambda}_a h_i^{(2)} 
\label{match}
\ea
with $Tr_{{\cal R}_i} T^a T^b={\rm l}_2({\cal R}^{a}_i)\delta^{ab}$ and  
$\hat{T}^{i\Lambda}=(2^{-3}B^{i\Lambda}_{b} n^{b} +
{1\over 4}\Gamma^{i\Lambda})$ 
the tadpole associated to the character ``$i\Lambda$'' and localized 
at one of the $4$ O7-planes. 
$t$-independence of the building blocks (\ref{spinsum}) implies 
exact agreement between the two expressions in (\ref{match})
\be
\beta^{(2)a}_{IR}=\beta^{(2)a}_{UV}
\label{bb}
\ee
Indeed plugging the contribution of any ${\cal N} =2$ supermultiplet 
in (\ref{betaD}), 
one can easily realize that ${h}_i^{(2)}$ numerically coincides 
with the Witten index associated to the character ``$i$''
of the internal $c=6$ SCFT theory, \ie $h_{\rm vector}^{(2)}=-2$
($h_{\rm hyper}^{(2)}=2$) for ${\cal N} ={2}$ vector (hyper-) multiplets. 
This integer is, of course, invariant under any modular 
transformations (in the character basis $S_i^j{h}_j^{(2)}={h}_i^{(2)}$)

As in \cite{bm}, we are thus allowed to relate direct and transverse
channel quantities through the algebraic
relations 
\begin{eqnarray} 
h^{(2)}_{i} A^{i\Lambda}_{ab} &=& 
h^{(2)}_{i} B^{i\Lambda}_{a}B^{i\Lambda}_{b}\nonumber\\
h^{(2)}_{i} M^{i\Lambda}_{a} &=& 
h^{(2)}_{i} B^{i\Lambda}_{a}\Gamma^{i\Lambda} \quad ,
\end{eqnarray}
which, once inserted in (\ref{match}), lead to a precise matching between IR 
and UV expressions (\ref{bb}). 

We conclude that RG-flows induced by ${\cal N}=2$ sectors of 
the gauge theory
living on the probe can be accurately described by the 
linear combination (\ref{match}) of massless tadpoles (one-point
functions) $\hat{T}^{i\Lambda}$ 
in the supergravity theory.
Notice that ${\cal N}=4$ sectors do not contributes to $h_i^{(2)}$ and 
therefore all relevant contributions share the lattice sum $W_2$.

Two comments are in order. First we would like to stress that in 
establishing the equivalence between $\beta$-functions 
and tadpoles for brane configurations satisfying the above 
requirement no large $N$ limit is required. 
Secondly, as we will see in details in the following, 
the supersymmetry requirement above allows for ${\cal N}=0,1$ 
brane configurations whose open string excitations can be assembled 
into sectors that    
realize different ${\cal N}=2$ superalgebras which intersect in
an effective less (or non-) supersymmetric gauge theory. Typical
examples are D3-branes probing a background of intersecting
D7-branes corresponding to F-theory at varying (but weak) coupling. 
Although the whole effective gauge theory in this case
only displays ${\cal N}=0,1$ supersymmetry, each sector 
of the free open string spectrum
enjoys ${\cal N}=2$ supersymmetry which contains the
${\cal N}=0,1$ intersection algebra. The cases studied
in \cite{aa} provide with
${\cal N}=0$ examples that fulfill all the above requirements. 
It would be interesting to test higher-loop corrections 
to the IR/UV correspondence proposed here for
these less supersymmetric brane configurations. 
In the ${\cal N}=1$ examples under study here one would
expect that a similar agreement should persist at least at the first 
subleading $1/N$ correction to each genus $g$ result.   
At this order the dominant contributions to the open string amplitudes 
come from surfaces with no handles and $g+1$ 
boundaries, $g$ of which are mapped onto the probe D3-branes and
only one on the background D7-branes. These give rise to effective ${\cal N}=2$ 
contributions that should not correct the one-loop result.
Amplitudes corresponding to surfaces with all the $g+1$ boundaries mapped onto 
the D3-branes should not contribute to the running because they are 
associated to ${\cal N}=4$ sectors of the open string spectrum 
\cite{vafaetal}.

Finally it is instructive to compare the NS-NS tadpoles/RG-flows
correspondence under discussion here with the 
IR/UV correspondence between R-R tadpoles and chiral gauge and 
gravitational anomalies
\cite{bm,bs,cp,auiq}. Indeed
the results presented here can be seen as the supersymmetric
analogue of the R-R-tadpoles/anomalies correspondence
\cite{bm}. In that case
the topological nature of the interactions allows stronger
results irrespective of the presence of spacetime  
unbroken supersymmetries. Notice that although the results 
presented here (whenever such IR/UV 
correspondence exists) are very similar
to the anomaly/tadpole correspondence established in \cite{bm} 
the group theory factors weighting the string contributions are 
obviously different and therefore different are the 
linear combinations of tadpoles entering the two kinds of computations. 
In addition
the topological index $h_i^{({D/2})}$, numerically
equal to the Witten index in $D+2$ dimensions, 
receives non-trivial contribution 
in $D$ dimension from the NS sectors in contrast with its anomaly cousin.

\section{D3-branes probes in type II and type IB backgrounds}

In this section we discuss several realizations of the IR/UV 
correspondence between RG-flows and NS-NS tadpoles for the case
of four-dimensional gauge theories living on D3-brane probes
of orientifold backgrounds. 
We will always start by specifying, from the brane data, the  
open string partition functions for vanishing background, 
\ie for $\epsilon_{ab}=0$  
in (\ref{direct},\ref{transverse}), in the direct and transverse
channel. 
In order to simplify the expressions we prefer 
to display formulas in terms of the amplitudes $\rho_{gh}$ 
rather than in terms of characters.
After translating the resulting amplitudes 
into the character basis through (\ref{zncharacter}) 
one can check  in each case that the Chan-Paton assigments
satisfy the consistency conditions (\ref{worldsheet}), \ie 
that transverse amplitudes factorize as in (\ref{transverse})
and that the M\"obius amplitude is an allowed projection of the 
Annulus amplitude.
The gauge theory will be determined by combining the massless
contents ($q^0$-coefficient) of the various partition
functions
\ba
\rho_{00}(0)&=&V+6O-4S-4C+.....\nonumber\\
\rho_{0h}(0)&=&V+2O(1+2cos {2\pi h\over N})-2(S+C)(1+cos {2\pi h\over N})+...
\nonumber\\
\rho_{gh}(0)&=&2O-S-C+...
\label{massless}
\ea
where $O$, $V$, $S$ and $C$ represent massless four-dimensional
scalars, vectors, left- and right- handed spinors respectively.  

The $\beta$-function coefficients (\ref{match}) can be easily
read from (\ref{match}) with the boundary data ${\cal A}^{i\Lambda}_{ab}$,
${\cal M}^{i\Lambda}_{a}$, $B_a^i$, $\Gamma^i$
defined through (\ref{direct},\ref{transverse})
and $h_i^{(2)}$ given by the substitutions
\be
(\rho_{00},\rho_{0h},\rho_{gh})\rightarrow (0,-4 sin^2{\pi h\over 
N},1) \quad .
\ee
We will denote the $\beta$-function coefficient for the D3 brane 
gauge group by $\beta_N$ with
\be
\beta_N\equiv 2\,\beta_{IR,UV}^{({2})N}
\ee
with $\beta_{IR,UV}^{({2})N}$\footnote{For $U(N)$ gauge groups 
we have an additional identical contribution coming from 
$\beta_{IR,UV}^{({2})\bar{N}}$} given 
by (\ref{match}). The overall factor of 2 is choosen 
in such a way to reproduce the familiar $\beta$-function 
coefficients for gauge group generators normalized as 
${\rm Tr}_{\rm fund} Q^2={1\over 2}$. 

Finally we would like to give a more precise meaning to 
the linear combination (\ref{match}) of tadpoles generating
the RG-flows in the four dimensional gauge theories   
we will present in the following. As we will see non-trivial
contributions to the closed string divergences only come 
from the $Z_2$-string amplitudes $\rho_{01}(0)=V+2O-4O+{\rm massive}$.
We see in particular that no contribution from RR states arise
as expected since these states do not couple to ${\rm Tr} {\cal F}^2$.  
Moreover noticing that the $h^{(2)}$ $\beta$-function coefficients that
weighted the string amplitude $\rho_{01}(0)$ satisfies
$h^{(2)}_{V+2O}=-h^{(2)}_{4O}$ one
can identified the linear combination in (\ref{match}) with the tadpole
of the ten dimensional dilaton $V+6O$. Our results then provide
a precise dictionary between this tadpole and $\beta$-function 
coefficients in the D3-brane gauge theory.

\subsection{D3-branes probing type II D7-backgrounds}

Let us start by considering $N$ D3-branes spanning the $0123$ directions
as a probe of a background geometry constructed out of two sets
of  $n_A$, $n_B$ intersecting D7-branes spanning,
besides the spacetime directions common to the D3-branes, the 
$A=(6789)$ and $B=(4589)$ hyperplanes respectively. 
We will also introduce the notation
$C=(4567)$ for the D7$_{A}$-D7$_{B}$ ND plane. 
The open string partition function in the direct channel 
is given by the integral $\int {d t\over t}$ of
\ba
{\cal A}&=&t^{-2}\left[\rho_{00}(N\bar{N} W_a W_b W_c
+ n_A\bar{n}_A W_a P_b P_c+n_B \bar{n}_B P_a W_b P_c\right.\nonumber\\
&&\left.
+\rho_{A0}(N \bar{n}_A+n_A\bar{N})W_a+\rho_{B0}
(N \bar{n}_B+n_B\bar{N})W_b
\right.\nonumber\\&&\left.
+\rho_{C0}(n_B \bar{n}_A+n_A\bar{n}_B)P_c\right]({it\over 2})
\label{d1}
\ea
where $a,b,c$ stand for the three planes $(45)$, $(67)$
and $(89)$. In the transverse channel one has
$\int {d \ell}$ of
\ba
\widetilde{\cal A}&=&2^{-5}\left[\rho_{00}(N\bar{N} 
{P_a P_b P_c\over v_a v_b v_c}
+n_A\bar{n}_A P_a W_b W_c{v_b v_c\over v_a} +
n_B\bar{n}_B W_a P_b W_c{v_a v_c\over v_b}\right.\nonumber\\
&&\left.+\rho_{0A}(N \bar{n}_A+n_A\bar{N}) {P_a\over v_a}
+\rho_{0B}(N \bar{n}_B+n_B\bar{N}){P_b\over v_b}
\right.\nonumber\\&&\left.
+\rho_{OC}(n_B \bar{n}_A+n_A\bar{n}_B) W_c v_c\right]({i\ell})
\label{t1}
\ea
Expressing (\ref{d1}) in terms of the $Z_2$-characters (\ref{zncharacter}), 
one can easily check that (\ref{t1}) factorizes properly at the origin
of the lattice sum and
therefore (\ref{d1}) defines a sensible CP-charge assignment.

Using the expansion (\ref{massless}) one can read the massless spectrum\\\\
\begin{tabular}{ll}
states & $U(N)\times U(n_A)\times U(n_B)$\\
 & \\
$V-S-C$  & $({\bf N\bar{N}},{\bf 1},{\bf 1})+
({\bf 1},{\bf n_A\bar{n}_A},{\bf 1})+({\bf 1},{\bf 1},{\bf n_B\bar{n}_B})$ \\
$3(2O-S-C)$ &  $({\bf N\bar{N}},{\bf 1},{\bf 1})+
({\bf 1},{\bf n_A\bar{n}_A},{\bf 1})+({\bf 1},{\bf 1},{\bf n_B\bar{n}_B})$ \\
$(2O-S-C)$& $({\bf N},{\bf \bar{n}_A},{\bf 1})+
({\bf N},{\bf 1},{\bf \bar{n}_B})
+({\bf 1},{\bf n_A},{\bf \bar{n}_B})+{\rm h.c.}$\\
\end{tabular}\\\\
The $SU(N)$ $\beta$-function coefficient
\be
\beta_N=(n_A+n_B)
\label{b1}
\ee
is always positive, thus leading to an IR free theory on the brane.
By setting $n_A=0$ or equivalently $n_B=0$ one gets a consistent 
theory with ${\cal N}=2$ supersymmetry.
The coefficient $\beta_N$ can be computed using
either (\ref{d1}) or (\ref{t1}) with the same result, \ie (\ref{b1}). 

\subsection{D3-branes in type IB backgrounds}

The next example we will consider is a stack of $N$ D3-branes 
exploring an $SO(n_A)$ singularity \cite{bds}. 
This background geometry is related to the
T-dual version (type IB or simply F-theory at constant coupling \cite{senf}) 
of type I theory on a $T^2$ torus with Wilson 
lines breaking the CP group to $SO(8)^4$ \cite{bps}. 
In the case at hand we 
consider moving
$(n_A-8)$ D7-branes far away from one of the $4$ O7-planes.
The D3-brane theory probes the geometry near this O7-plane.
A sensible assigment of CP-charges leads to the open
string partition function
\ba
{\cal A}&=&{t^{-2}\over 2}\left[\rho_{00}(N^2 W_a W_b W_c
+ n_A^2 W_a P_b P_c) +2 \rho_{A0} N n_A W_a\right]\left({it\over 2}\right)
\nonumber\\
{\cal M}&=&-{t^{-2}\over 2}(\rho_{00} n_A W_a P_b P_c
-\rho_{0A} N W_a)\left({it\over 2}+{1\over 2}\right)
\label{d2}
\ea
Rewriting (\ref{d2}) in terms of the closed string variables one has 
($\int {d \ell}$)
\ba
\widetilde{\cal A}&=&{2^{-5}\over 2}\left[\rho_{00}
(N^2 {P_a P_b P_c\over v_a v_b v_c}
+n_A^2 P_a W_b W_c{v_b v_c\over v_a})+2\rho_{0A} N n_A {P_a\over v_a}
\right]({i\ell})\nonumber\\
\widetilde{\cal M}&=&-{2 \over 2}\left[\rho_{00} n_A W^{\rm e}_a W^{\rm e}_b 
P^{\rm e}_c{v_a v_b\over v_c} 
+\rho_{0A} N {P^{\rm e}_a\over v_a}\right]({i\ell})
\ea
As expected,   
the transverse amplitudes factorize properly.

The massless spectrum consists in\\\\
\begin{tabular}{ll}
states & $Sp(N)\times SO(n_A)$\\
 & \\
$V-S-C$  & $({\bf {1\over 2}N(N+1)},{\bf 1})+
({\bf 1},{\bf  {1 \over 2} n_A(n_A-1)})$\\
$(2O-S-C)$  & $({\bf {1\over 2}N(N+1)},{\bf 1})+
({\bf 1},{\bf  {1\over 2}n_A(n_A-1)})$\\
$2(2O-S-C)$ &  $({\bf {1\over 2}N(N-1)},{\bf 1})+
({\bf 1},{\bf  {1\over 2}n_A(n_A-1)})$\\
$(2O-S-C)$  & $({\bf N},{\bf n_A})$\\
\end{tabular}\\\\
The one-loop $\beta$-function coefficient for the $Sp(N)$ gauge coupling 
\be
\beta_N={1\over 2}(n_A-8)
\label{b2}
\ee
can be made
negative (for $n_A < 8$) thanks to the presence of the O7-planes.
We thus expect the corresponding theory on the probe to be non-trivial 
in the IR.

\subsection{D3-branes probes of a $C^4/Z_2$ Orientifold singularity}

We now start from a T-dual version of the $T^4/Z_2$ orientifold
of type IIB \cite{bs, ps, gp} where D9-D5 branes become D3-D7 branes
and we send as before the radius of all internal directions to infinity
where tadpole conditions can be relaxed. The open string partition
function is given by
($\int {d t\over t}$)
\ba
{\cal A}&=&{t^{-2} \over 4}\left[\rho_{00}((N+\bar{N})^2 W_a W_b W_c
+(n_A+\bar{n}_A)^2 W_a P_b P_c)\right.\nonumber\\
&&\left.
-\rho_{0A}((N-\bar{N})^2+(n_A-\bar{n}_A)^2)W_a
+2\rho_{A0}(N+\bar{N})(n_A+\bar{n}_A)W_a
\right.\nonumber\\ && \left.
-2\rho_{AA}(N-\bar{N})(n_A-\bar{n}_A)W_a\right]\left({it\over 2}\right)
\nonumber\\
{\cal M}&=&-{t^{-2} \over 4}\left[\rho_{00}((N+\bar{N}) W_a W_b W_c
+(n_A+\bar{n}_A) W_a P_b P_c)\right.\nonumber\\
&&\left. -\rho_{0A}(N+\bar{N}+n_A+\bar{n}_A)W_a
\right]\left({it\over 2}+{1\over 2}\right)
\ea
Rewritten in terms of the closed string variables one has 
($\int {d \ell}$)
\ba
\widetilde{A}&=&{2^{-5}\over 4}\left[\rho_{00}((N+\bar{N})^2 
{P_a P_b P_c\over v_a v_b v_c}
+(n_A+\bar{n}_A)^2 W_a W_b P_c{v_a v_b\over v_c})+
\right.\nonumber\\&&\left.
2\rho_{0A} (N+\bar{N})(n_A+\bar{n}_A){P_a\over v_a}
-16\rho_{A0}((N-\bar{N})^2
+(n_A-\bar{n}_A)^2){P_a\over v_a}\right.\nonumber\\&&\left.
+8\rho_{AA}(N-\bar{N})(n_A-\bar{n}_A){P_a\over v_a}\right]({i\ell})\nonumber\\
\widetilde{M}&=&-{2 \over 4}\left[\rho_{00}
(((N+\bar{N}){P^{\rm e}_a P^{\rm e}_b P^{\rm e}_c\over v_a v_b v_c}
+(n_A+\bar{n}_A)P^{\rm e}_a W^{\rm e}_b W^{\rm e}_c{v_b v_c\over v_a})\right.\nonumber\\
&&\left.+\rho_{0A}(N+\bar{N}+n_A+\bar{n}_A){P^{\rm e}_a\over 
v_a}\right]({i\ell})
\ea 
The massless spectrum is given by\\\\
\begin{tabular}{ll}
states & $U(N)\times U(n_A)$\\
 & \\
$V-S-C$  & $({\bf N\bar{N}},{\bf 1})+
({\bf 1},{\bf n_A\bar{n}_A})$ \\
$(2O-S-C)$  & $({\bf N\bar{N}},{\bf 1})+
({\bf 1},{\bf n_A\bar{n}_A})$ \\
$2(2O-S-C)$ &  $({\bf {1\over 2}N(N-1)},{\bf 1})+
({\bf 1},{\bf  {1\over 2}n_A(n_A-1)}) + h.c.$\\
$(2O-S-C)$& $({\bf N},{\bf \bar{n}_A})+{\rm h.c.}$\\
\end{tabular}\\\\
The $SU(N)$ $\beta$-function coefficient at one-loop 
\be
\beta_N={1\over 2}(n_A+\bar{n}_A-8) 
\label{b3}
\ee
receives a negative contribution from 
the orientifold planes in the closed string description. 
Notice that a conformal point $n_A=\bar{n}_A=4$ exists 
associated to a democratic distribution of the sixteen pairs
of D7-branes between the four O7-planes.

\subsection{D3-branes probes of a $C^6/Z_2\times Z_2$ Orientifold singularity}
 
Finally we will consider $N$ D3 branes in 
a $T^6$ orientifold generated by the group elements $\Omega,Z_2^A,Z_2^B$.
This is just a T-dual version of the $N=1$ type I background
studied in \cite{bl}. In addition to the 
D3-branes and O3-planes, spanning 
the directions $0123$ common to all the branes, three groups of 
D7-branes and O7-planes are require to invade
the $A=(6789)$, $B=(4589)$ and $C=(4567)$ hyperplanes respectively. 
String amplitudes $\rho_{gh}$ and $Z_2\times Z_2$ characters
are related to each other by 
\be
\pmatrix{\chi_{g0}\cr \chi_{gA}\cr \chi_{gB}\cr \chi_{gC}}
=S_{SO(8)}\pmatrix{\rho_{g0}\Sigma_{g0}
\cr \rho_{gA}\Sigma_{gA}\cr \rho_{gB}\Sigma_{gB}\cr \rho_{gC}\Sigma_{gC}}
={1\over 2}\pmatrix{+& +& +& +\cr +& +& -& -\cr +& -& -& +\cr +& -& +& -\cr} 
\pmatrix{\rho_{g0}\Sigma_{g0}\cr \rho_{gA}\Sigma_{gA}\cr 
\rho_{gB}\Sigma_{gB}\cr \rho_{gC}\Sigma_{gC}}
\label{sso8}
\ee
which generalizes (\ref{zncharacter}) to this case. 
The open string partition function in the direct channel reads \cite{bl} 
($\int {d t\over t}$)
\ba
{\cal A}&=&{t^{-2} \over 2}\left[\rho_{00}(N^2 W_a W_b W_c
+n_A^2 W_a P_b P_c+n_B^2 P_a W_b P_c+n_C^2 P_a P_b W_c)\right.\nonumber\\
&&\left.
+2\rho_{A0}(N n_A W_a+n_B n_C P_a)+2\rho_{B0}(N n_B W_b+n_A n_C P_b)
\right.\nonumber\\&&\left.
+2\rho_{C0}(N n_C W_c+n_A n_B P_c)\right]\left({it\over 2}\right)\nonumber\\ 
{\cal M}&=&{t^{-2} \over 2}\left[-{1\over 2}\rho_{00}(N W_a W_b W_c
+n_A W_a P_b P_c+n_B P_a W_b P_c+n_C P_a P_b W_c)\right.\nonumber\\
&&\left.
+{1\over 2}\rho_{0A}((N+n_A)W_a+(n_B+n_C) P_a)
+{1\over 2}\rho_{0B}((N+n_B)W_b+(n_A+ n_C) P_b)
\right.\nonumber\\&&\left.
+{1\over 2}\rho_{0C}((N+n_C)W_c+(n_A+n_B) P_c)
\right]\left({it\over 2}+{1\over 2}\right)
\label{dbl}
\ea
Once rewritten in terms of the closed string variables,  (\ref{dbl})
yields ($\int {d\ell}$)
\ba
\widetilde{\cal A}&=&{2^{-4}\over 4}\left[\rho_{00}(N^2 
{P_a P_b P_c\over v_a v_b v_c}
+n_A^2 P_a W_b W_c{v_b v_c\over v_a} +
n_B^2 W_a P_b W_c{v_a v_c\over v_b}+
n_C^2 W_a W_b P_c{v_a v_b\over v_c})\right.\nonumber\\
&&\left.+2\rho_{0A}(N n_A {P_a\over v_a}+n_B n_C W_a v_a)
+2\rho_{0B}(N n_B{P_b\over v_b}+ n_A n_C W_b v_b)
\right.\nonumber\\&&\left.
+2\rho_{OC}(N n_C {P_c\over v_c}+n_A n_B W_c v_c)\right]({i\ell})\nonumber\\
\widetilde{\cal M}&=&-{2 \over 4}\left[\rho_{00}(
N{P^{\rm e}_a P^{\rm e}_b P^{\rm e}_c\over v_a v_b v_c}
+n_A P^{\rm e}_a W^{\rm e}_b W^{\rm e}_c{v_c v_b\over v_a} +
n_B W^{\rm e}_a P^{\rm e}_b W^{\rm e}_c{v_a v_c\over v_b}+
n_C W^{\rm e}_a W^{\rm e}_b P^{\rm e}_c{v_b v_a\over v_c})\right.\nonumber\\
&&\left.+\rho_{0A}((N+n_A){P^{\rm e}_a\over v_a}+(n_B+n_C)W^{\rm e}_a v_a)
+\rho_{0B}((N+n_B){P^{\rm e}_b\over v_b}+(n_A+n_C)W^{\rm e}_b v_b)
\right.\nonumber\\&&\left.
+\rho_{OC}((N+n_C){P^{\rm e}_c\over v_c}+(n_A+n_B) W^{\rm e}_c v_c)
\right]({i\ell})
\ea
Once again, one can check that 
the transverse amplitudes factorize properly. In particular for the 
states with vanishing winding and momentum one finds a sum of perfect 
squares for $\widetilde{\cal A}$ and a compatible sum of products for 
$\widetilde{\cal M}$.

The massless spectrum reads\\\\
\begin{tabular}{ll}
states & $Sp(N)\times Sp(n_A)\times Sp(n_B)\times Sp(n_C)$\\
 & \\
$V-S-C$  & $({\bf {1\over 2}N(N+1)},{\bf 1},{\bf 1},{\bf 1})+
({\bf 1},{\bf  {1\over 2}n_A(n_A+1)},{\bf 1},{\bf 1})$\\
 &$({\bf 1},{\bf 1},{\bf  {1\over 2}n_B(n_B+1)},{\bf 1})
+({\bf 1},{\bf 1},{\bf 1},{\bf  {1\over 2}n_C(n_C+1)})$\\
$3(2O-S-C)$  & $({\bf {1\over 2}N(N-1)},{\bf 1},{\bf 1},{\bf 1})+
({\bf 1},{\bf  {1\over 2}n_A(n_A-1)},{\bf 1},{\bf 1})$\\
 &$({\bf 1},{\bf 1},{\bf  {1\over 2}n_B(n_B-1)},{\bf 1})
+({\bf 1},{\bf 1},{\bf 1},{\bf  {1\over 2}n_C(n_C-1)})$\\ 
$(2O-S-C)$  & $({\bf N},{\bf n_A},{\bf 1},{\bf 1})+
({\bf N},{\bf 1},{\bf n_B},{\bf 1})+({\bf N},{\bf 1},{\bf 1},{\bf n_C})$\\
 &$ ({\bf 1},{\bf n_A},{\bf n_B},{\bf 1})+
({\bf 1},{\bf n_A},{\bf 1},{\bf n_C})+({\bf 1},{\bf 1},{\bf n_B},{\bf n_C})$\\
\end{tabular}\\\\
Depending on the number of background D7-branes of various kinds,
the $Sp(N)$ $\beta$-function coefficient
\be
\beta_N={1\over 2}(n_A+n_B+n_C-12)
\label{b4}
\ee
can be vanishing, positive or negative.
A (super)conformal four dimensional gauge theory is found for 
$n_A+n_B+n_C=12$. 

\section{AdS/CFT correspondence and final comments}

Soon after Maldacena's proposal of identifying type IIB superstring on
$AdS_5 \times S^{5}$ with $N$ R-R 5-form fluxes and
${\cal N} = 4$ SYM with gauge group $SU(N)$ \cite{jm}, many other candidate 
supergravity duals of superconformal 
field theories in $D=4$ have been proposed \cite{magoo}.
In particular D3-branes exploring F-theory backgrounds at constant 
coupling were identified as a natural arena for the holographic study 
of 
superconformal field theories \cite{afm}. The near horizon geometry of 
a stack of $N$ D3-branes in an F-theory background at constant imaginary 
coupling is given by 
\ba
ds^{2} &&= H(y)^{-1/2} \eta_{\mu\nu} dx^{\mu} dx^{\nu} + 
H(y)^{1/2} g_{ij} dy^{i} dy^{j} \nonumber \\
e^{\phi} &&= g_{s} \nonumber \\
F_{5} &&= {N} (\epsilon_{5} + * \epsilon_{5})  \quad .
\label{d3d7solution}
\ea
It locally looks like $AdS_{5}\times E_{5}$,  where $E_{5}$ is 
some Einstein manifold 
that replaces the familiar $S^{5}$ \cite{klebwit, kehag}. The scale of the cosmological 
constants of the two factors is set by $L^{4} = g_{s}N (\alpr)^{2} $, 
$\epsilon_{5}$ is the $AdS_{5}$ volume form and 
$H$ is a harmonic function in the 
Ricci flat metric $g_{ij}$ transverse to the 3-branes.
All other fields are set to zero in the background.

We are interested in the cases when the mutually non-local 
7-branes are grouped in such a way as to give rise to 
bound-states of D7-branes and O7-planes at constant weak coupling 
\cite{senf}. These configurations admit a perturbative description in 
terms of open and unoriented strings. The metric of the transverse
space is explicitly known to be \cite{afm}
\be
dr^{2} + r^{2} ds_{5}^{2} = \sum_{a=1}^{3} {|dz_{a}|^{2} \over 
|z_{a}|^{\alpha_{a}} }
\label{einstein}
\ee
with $r^{2} = \sum_{a=1}^{3} |w_{a}|^{2}$ and 
$w_{a} = z_{a}^{1-{\alpha_{a}\over 2}}$. The exponent $\alpha_{a}$ is 
1 (0) when there are (no) 7-branes invading the complex plane labeled by $a$.

After performing two T-dualities along directions transverse to the 
3-branes, the above F-theory background 
is mapped to a perturbative type I background. 
In the previous sections, we have shown that the 
factorization of unoriented/open string vacuum amplitudes in the 
presence of background gauge fields allows one to deduce the coupling 
of the bulk closed string states and in principle to reconstruct the 
metric, dilaton and p-form backgrounds.

Following \cite{aa} and relying on the results of the previous sections,
we now turn to consider
cases in which the dilaton varies in the F-theory background.
Indeed by displacing some of the D7-branes from the orientifold planes
it is possible to generate a local tadpole for the 10-D dilaton
on the disk and the projective plane. In the large $N$ counting these 
contributions are subleading and are holographically dual 
to corrections that drive a logarithmic running of
the gauge coupling in theories that are (super)conformal in the large 
$N$ limit. The expansion in powers of $1/N$ is systematic and the 
coupled system of equations admits a solution that is asymptotic to
$AdS_{5}\times E_{5}$.

The relevant system of equations has been studied by \cite{aa}. We will 
perform a similar first-order analysis for more general F-theory-like
configurations with ``dilaton'' tadpoles. Since the R-R axion 
and the complex antisymmetric tensor appear at least quadratically 
in the equations of motion for the metric, dilaton and four-index 
antisymmetric tensor it is consistent for our purposes 
to truncate the theory to
the latter set of fields.
In the presence of tadpoles, the relevant equations of motion read
\ba
e^{-2\phi} && \left[ 8 (\nabla\phi)^{2} - 8 \nabla^{2}\phi - 2 R 
\right] = 
e^{-\phi} T^{\phi} + \ldots \nonumber \\
e^{-2\phi} && \left[ R_{MN} + 2 \nabla_{M}\nabla_{N}\phi \right] - 
{1\over 96} \left[ F_{MPQRS}F_{N}{}^{PQRS} 
-{1\over 10}g_{MN} F_{PQRST}F^{PQRST}
\right] \nonumber\\
&& = 
e^{-\phi} T^{g}_{MN} + \ldots \nonumber \\
\nabla^{M}&&F_{MNPQR} = 
 T^{A}_{NPQR} - \ldots  \quad .  
\ea
Notice the absence of the suppression factor $e^{+\phi_{o}}= 
g_{s}\approx 1/N$ in front of the RR-tadpole in the last equation. 
The $1/N$ suppression comes in this case from the overall scaling of 
the background self-dual five-form with $N$.

Starting from the solution (\ref{d3d7solution}) of the 
above equations of motion
at vanishing tadpoles, one can 
turn on the tadpoles and linearize the system around (\ref{d3d7solution}).
An ansatz for the metric and the 10-D dilaton that is compatible with  
four-dimensional Poincar\'e invariance is
\ba
ds^{2} &&=  d\tau^{2} + e^{2\lambda ( \tau )}\eta_{\mu\nu} dx^{\mu} dx^{\nu} + 
 e^{2\nu ( \tau )} \hat{g}_{\alpha\beta}d\theta^{\alpha} d\theta^{\beta}\\
\phi &&= \phi (\tau) 
\label{ansatz}
\ea
where $\hat{g}_{\alpha\beta}$ is the constant curvature metric of the Einstein 
space $E_{5}$ defined by (\ref{einstein}).
To the order at which we are going to work the correction to the R-R 
fields, including the four-form potential, will play no role 
in the evolution of the dilaton.
This can be easily seen using the traceless property of
the stress energy tensor associated to a four form RR-potential
in order to eliminate the curvature scalar from the dilaton 
equation. Indeed barring the 
fields with vanishing background, the scalar curvature is
entirely given in terms of the dilaton by 
\be
R = - 2 \nabla^{2}\phi - {5\over 2} T^{\phi} e^{+\phi}
\label{elementarywatson}
\ee
Plugging $R$ into the equation of motion for the dilaton 
yields
\be
8 (\nabla\phi)^{2} - 4 \nabla^{2}\phi = 4 T^{\phi} e^{+\phi} 
\label{verysimple}
\ee
that combined with the ansatz (\ref{ansatz}) for the dilaton and the metric leads 
to
\be
\ddot\phi + (4 \dot\lambda + 5 
\dot\nu) \dot\phi - 2 (\dot\phi)^{2} = T^{\phi} e^{+\phi} 
\label{evensimpler}
\ee
where a dot denotes a derivative with respect to the radial transverse
coordinate $\tau$.
Eq. (\ref{evensimpler}) agrees with the corresponding equation derived in \cite{aa} 
for the simpler cases with vanishing R-R 
tadpoles up to the overall factor in the source term.

Expanding with respect to the solution with $e^{\phi_{o}}= g_{s}$,
$\lambda_{o}= \tau/L$ and $e^{\nu_{o}}= L^{2}$ for vanishing tadpoles  
\ba
\phi &&= \phi_{o} + {1\over N} \phi_{1} + \ldots \\
\lambda &&= \lambda_{o} + {1\over N} \lambda_{1} + \ldots \\
\nu &&= \nu_{o}+ {1\over N} \nu_{1} + \ldots  
\ea
one immediately sees that $\phi_{1}$ decouples from the remaining 
fields $\lambda_{1}$ and $\nu_{1}$, since $\dot\phi_{o} = 0$.
Using $\dot\lambda_{o} = 1/L$ we are left with
\be
\ddot\phi_{1} + {4 \over L} \dot\phi_{1} = 4 T^{\phi}   
\label{supersimple}
\ee
that admits $\phi_{1} = T_{\phi} L \tau $ as an obvious solution.
After a proper definition of the radial 
variable $u = \int e^{\lambda (\tau)} d\tau \approx L e^{\tau/L}$ 
the correction to the dilaton turns out to be logarithmic in $u$ 
with $\beta$-function coefficient $T_\phi$ as in the case
studied by \cite{aa}. Notice that it is the 10-D dilaton $\phi$ that has to be 
holographically identified with the gauge coupling \cite{pw}.

We interpret this as a logarithmic running of the coupling constant 
in the boundary gauge theory. Notice that our argument above has the 
virtue to 
establishing a precise identification of the one-loop $\beta$-function
coefficient with the 10-D dilaton tadpole. Being the dilaton tadpole 
associated to a  
UV divergence we expect it to be independent of the large distance 
properties of the background and we find no need to perform
a fullfledged string computation around $AdS_{5}$ 
but rather use the flat-spacetime 
results of the previous sections.

In the cases that we have explicitly studied the coefficient can be 
negative only when orientifold planes are present: brane 
configurations tend to have positive $\beta$-function and thus to expose a 
trivial
IR behaviour. We do not know whether this is a serious drawback for 
the purposes of describing non super-conformal gauge theories 
in terms of brane probes or simply 
a limitation imposed by the configurations we have considered.

Moreover we also don't know whether two- and higher loop 
contributions can spoil the picture especially when supersymmetry is 
broken and the scalar self-couplings tend to run in the wrong direction.
Clearly further study is needed.
Extracting the $\beta$-function coefficient from a genuine two-loop 
computation in theories of open and unoriented strings seems feasable 
and we hope to return to this issue in the near future.

Although higher loops might show a systematic pattern \cite{kv, dbvv} 
the world-sheet duality between open and closed string channels 
gets more involve. 
Nevertheless it can prove to be an 
unprecedented tool in the interplay between gauge theory and gravity:
a bridge over trouble waters.

\vspace*{0.3cm}

{\large{\bf Acknowledgements}}

\vspace*{0.3cm}
\noindent
We would like to acknowledge fruitful discussions with 
C.~Angelantonj, A.~Armoni, D.~Freedman, M.~Green, S.~Kovacs, 
J.~Minahan, G.~Pradisi, G.~Rossi, A.~Sagnotti and
Ya.~Stanev. This work was in part supported by the EEC
contract HPRN-CT-2000-00122 and the INTAS project 991590.

\end{document}